\documentclass[aps,pra,superscriptaddress,twocolumn, floatfix]{revtex4-1}
\usepackage[
  bookmarks=true,
  colorlinks,
  linkcolor=blue,
  urlcolor=blue,
  citecolor=blue,
  plainpages=false,
  pdfpagelabels,
  final,
  breaklinks=true
]{hyperref}

\usepackage[utf8]{inputenc}
\usepackage{braket}
\usepackage{amssymb}
\usepackage{amsmath}
\usepackage{dsfont}
\usepackage{bm}
\usepackage{graphicx}
\usepackage{subfigure}
\usepackage{color}

\usepackage{times}

\definecolor{fuchsia}{rgb}{1.0, 0.0, 1.0}

\newcommand{\Bsigma}{\boldsymbol{\sigma}}
\newcommand{\Btheta}{\boldsymbol{\theta}}
\newcommand{\Bomega}{\boldsymbol{\omega}}
\newcommand{\Plambda}{\lambda_{\Btheta}(\Bsigma)}

\DeclareRobustCommand{\diamondtimes}{%
  \mathbin{\text{\rotatebox[origin=c]{45}{$\boxplus$}}}%
}

\begin{document}

\title{Entanglement Forging with generative neural network models}

\author{Patrick Huembeli}
\email[]{patrick.huembeli$\diamondtimes$menten.ai}
\affiliation{Menten AI, Inc., Palo Alto, California, United States of America}
\affiliation{Institute of Physics, \'Ecole Polytechnique F\'ed\'erale de Lausanne (EPFL), CH-1015 Lausanne, Switzerland}


\author{Giuseppe Carleo}
\affiliation{Institute of Physics, \'Ecole Polytechnique F\'ed\'erale de Lausanne (EPFL), CH-1015 Lausanne, Switzerland}

\author{Antonio Mezzacapo}
\affiliation{IBM Quantum, IBM T. J. Watson Research Center, Yorktown Heights, New York 10598, USA}

\date{\today}

\begin{abstract}
The optimal use of quantum and classical computational techniques together is important to address problems that cannot be easily solved by quantum computations alone. This is the case of the ground state problem for quantum many-body systems. 
We show here that probabilistic generative models can work in conjunction with quantum algorithms to design hybrid quantum-classical variational ans\"atze that forge entanglement to lower quantum resource overhead. The variational ans\"atze comprise parametrized quantum circuits on two separate quantum registers, and a classical generative neural network that can entangle them by learning a Schmidt decomposition of the whole system. The method presented is efficient in terms of the number of measurements required to achieve fixed precision on expected values of observables. To demonstrate its effectiveness, we perform numerical experiments on the transverse field Ising model in one and two dimensions, and fermionic systems such as the t-V Hamiltonian of spinless fermions on a lattice.
\end{abstract}
\maketitle

\section{Introduction}
In recent years quantum computing processors have been steadily improved and some small-scale simulations of quantum chemistry ~\cite{kandala2017hardware, kandala2019error, mccaskey2019quantum, google2020hartree} and many-body~\cite{zhukov2018algorithmic, smith2019simulating} simulations have been successfully executed on quantum processors. It is believed that for ground state problems quantum computers have an advantage over their classical counterpart, however no quantum algorithm that can solve it exactly in the most general setting exists~\cite{kempe2006complexity}, especially when taking into account finite quantum resources. At the same time, machine learning (ML) algorithms and especially generative neural networks (NNs) have shown to be efficient tools to find ground state energies of many-body systems in the form of neural network quantum states~\cite{carleo2017solving}. 
There are several different proposals to classically correlate observables from non-interacting systems \cite{ayralQuantumDivideCompute2020, ayralQuantumDivideCompute2021,  bauerHybridQuantumClassicalApproach2016, bravyiComplexityQuantumImpurity2017, bravyiTradingClassicalQuantum2016, eddinsDoublingSizeQuantum2022, kawashimaOptimizingElectronicStructure2021, kreulaFewqubitQuantumclassicalSimulation2016, mitaraiConstructingVirtualTwoqubit2021, pengSimulatingLargeQuantum2020, smartQuantumSolverContracted2021, tangCutQCUsingSmall2021, yamazakiPracticalApplicationNearTerm2018, yuanQuantumSimulationHybrid2021a, marshall2022high} all with the purpose of reducing the amount of quantum resources to be used by a quantum processor. 

Classically forged entanglement as introduced in~\cite{eddinsDoublingSizeQuantum2022} refers to the emulation of properties of a $2N$-qubit state via the embedding of two $N$-qubit subsystems in a classical computation. To better understand how this classical computation can be achieved one can start from the Schmidt decomposition of a state
\begin{align}
    \ket{\psi} = U_A \otimes V_B \sum_{\Bsigma} \lambda_{\Bsigma} \ket{\Bsigma}_A \ket{\Bsigma}_B,
\end{align}
where $\Bsigma = (\sigma_1, \dots, \sigma_N) \in \{0,1\}^N$ denotes a $N$-bit string, $U_A$ and $V_B$ are unitaries acting on the subsystems $A$ and $B$ and $\lambda_{\Bsigma}$ are the Schmidt coefficients which are non-negative. The Schmidt decomposition is the most general form a two-partite pure state can be written in. If one manages to determine both the Schmidt coefficients $\lambda_{\Bsigma}$, and the unitaries $U_A$,$V_B$ exactly, one can theoretically forge any pure state. 
Inspired by the Schmidt decomposition we introduce a quantum-classical hybrid ansatz. We show that with the use of a generative NN model it is possible to approximate the Schmidt coefficients and, together with trainable circuit unitaries $U_A$, one can approximate the ground state of many-body states with high accuracy.
More concretely, we find the ground states of the transverse field Ising model (TFIM) with periodic boundary conditions in 1 and 2 dimensions at its critical point with a classically forged state. We show that the ground state energy is reached with high accuracy and that the $\langle Z_i Z_j \rangle$ correlators of the exact diagonalization can be reproduced. Furthermore, we show that our approach can also be applied to fermionic system such as the t-V model of spinless fermions and forge the required entanglement.

\section{Heisenberg Forging}
We start by summarizing the entanglement forging framework introduced in~\cite{eddinsDoublingSizeQuantum2022}, specializing it to
states where we set $U_A = V_B$. We can then drop the subscript for the unitary operator $U$, and obtain the state
\begin{align}
\label{forged_state}
    \ket{\psi} = U \otimes U \sum_{\Bsigma} \lambda_{\Bsigma} \ket{\Bsigma}_A \ket{\Bsigma}_B.
\end{align}
By parametrizing $U$ and $\lambda_{\Bsigma}$ we create an ansatz that is restricted to systems which are symmetric under the permutation of systems $A$ and $B$, such as translational invariant systems. We are interested to estimate observables such as for example Hamiltonians of the form $H = \sum_i h_i$ with $h_i \in \{I, X, Y, Z \}^{\otimes N}$, where $I, X, Y, Z$ are the Pauli matrices. We denote a general Pauli observable $O_A \otimes O_B$ via two operators that act on each subsystem respectively and the goal is to estimate the expectation value  
\begin{align}
    \mu = \braket{\psi \vert O_A \otimes O_B \vert \psi}.
\end{align}
The expectation value for an observable defined only on one subsystem (e.g. $O_B = I$)  simplifies to 
$\mu = \sum_{\Bsigma} \lambda_{\Bsigma}^2 \braket{\Bsigma \vert U^{\dag} O_A U \vert \Bsigma}$, which can be estimated via sampling $\Bsigma \sim \lambda_{\Bsigma}^2$, preparing the circuit $U \ket{\Bsigma}$ and measuring $\braket{\Bsigma \vert U^{\dag} O_A U \vert \Bsigma}$. To estimate an observable acting on both subsystems we can rewrite
\begin{align}
\label{Eq:extend_OA}
    &O_A \otimes O_B + O_B \otimes O_A = \frac{a_0}{2}\bigg( \{O_A, O_B\} \otimes I \nonumber \\ 
    &+ I \otimes \{O_A, O_B\}\bigg) 
    + \sum_{\alpha, \beta \in \{0,1\}} a_{\alpha, \beta} C_{\alpha, \beta}^* \otimes C_{\alpha, \beta} ,    
\end{align}
where $\{O_A,O_B\} = O_A O_B + O_B O_A$ denotes the anti-commutator, $\vert a_{\alpha, \beta}\vert \leq 1$ are real coefficients and $C_{\alpha, \beta}$ are $N$-qubit Clifford operators which we characterize below. We make use of the permutational invariance of the system that allows us to swap the observables of the subsystems and symmetrize $\mu$ for the swap of $O_A$ and $O_B$, $\mu = \left(\braket{\psi \vert O_A \otimes O_B \vert \psi} + \braket{\psi \vert O_B \otimes O_A \vert \psi}\right) / 2$. Considering equation~\ref{Eq:extend_OA}, we obtain
\begin{align}
\label{eq:Expect_off_diagonal}
    \mu &= a_0 \sum_{\Bsigma} \lambda_{\Bsigma}^2 \mathrm{Re}(\braket{\Bsigma \vert U^{\dag} O_AO_B U \vert \Bsigma }) \nonumber \\
    &+  \sum_{\alpha, \beta \in \{0,1\}} \frac{a_{\alpha, \beta}}{2} \braket{\psi \vert C_{\alpha, \beta}^* \otimes C_{\alpha, \beta} \vert \psi},
\end{align}
we use the short hand notation
\begin{align}
\label{eq:Forging_complete}
    \mu_{\alpha, \beta} &= \braket{\psi \vert C_{\alpha, \beta}^* \otimes C_{\alpha, \beta} \vert \psi} \nonumber \\
    &  = \sum_{\Bsigma, \Bsigma'} \lambda_{\Bsigma} \lambda_{\Bsigma'} \vert \braket{\Bsigma' \vert U^{\dag} C_{\alpha, \beta} U\vert \Bsigma } \vert^2 
\end{align}
For two commuting Pauli observables $O_A,~O_B \in \{I, X, Y, Z \}^{\otimes N}$ there exists an $N$-qubit Clifford circuit $V$ and a pair of qubits $p,~q \in \{1,2 \dots, N \}$ such that we can write $O_A = V Z_p V^{\dag}$ and $O_B = V Z_q V^{\dag}$. 
The aforementioned Clifford gates can be written as 
\begin{align}
    C_{\alpha, \beta} = V \frac{1}{2} \left( \mathds{1} + (-1)^{\alpha} Z_p + (-1)^{\beta} Z_q - (-1)^{\alpha + \beta} Z_p Z_q  \right) V^{\dag}.
\end{align}
The unitary $C_{\alpha, \beta}$ can also be expressed as a sequence of standard one- and two-qubit gates. If $p\neq q$ we can write 
$C_{\alpha, \beta} = V X_p^{\alpha} X_q^{\beta} \text{CZ}_{p, q} X_p^{\alpha} X_q^{\beta} V^{\dag}$ if $p=q$, $C_{\alpha, \beta} = V X_p^{\alpha} X_q^{\beta} P_p((\alpha + \beta - 1) \cdot \pi) X_p^{\alpha} X_q^{\beta} V^{\dag}$. $P_p$ is the phase gate applied to qubit $p$.
In the case of the TFIM Hamiltonian $V = \mathds{1}$.

The first term of equation~\ref{eq:Expect_off_diagonal} can be evaluated the same way as for observables that only act on one subsystem, because $O_A O_B$ does not denote a tensor product anymore but a simple multiplication of the two Pauli strings $O_A$ and $O_B$, acting on one of the subsystems. To evaluate the 2nd term of equation~\ref{eq:Expect_off_diagonal} we define the function $R(\Bsigma, \Bsigma') = \lambda_{\Bsigma'} /  \lambda_{\Bsigma}$ and interpret $ \vert \braket{\Bsigma' \vert U^{\dag} C_{\alpha, \beta} U\vert \Bsigma } \vert^2$ as a conditional probability $p_{\alpha, \beta}(\Bsigma' \vert \Bsigma)$ of how likely it is to sample $\Bsigma'$ from a circuit $U^{\dag} C_{\alpha, \beta} U \ket{\Bsigma }$. As a consequence, we can rewrite
\begin{align}
\label{eq:mu_alpha_beta}
    \mu_{\alpha, \beta}  = \sum_{\Bsigma} \lambda_{\Bsigma}^2 \sum_{\Bsigma'} R(\Bsigma, \Bsigma') p_{\alpha, \beta}(\Bsigma' \vert \Bsigma),
\end{align}
which can be estimated via sampling $\Bsigma \sim \lambda_{\Bsigma}^2$ and $\Bsigma' \sim p(\Bsigma'\vert \Bsigma)$. This simplification is a consequence of setting $U_A = V_B$ at the beginning of this section. Without this restriction of the ansatz, at least to our knowledge, it would not be possible to estimate $\mu_{\alpha, \beta}$ efficiently. The simplification also addresses some of the issues with the scalability of measurement requirements arising in other attempts to combine classical and quantum sampling, such as in~\cite{hugginsUnbiasingFermionicQuantum2021, zhangVariationalQuantumNeuralHybrid2021}.

\subsection{Neural Network Ansatz}
If we assume a small number of Schmidt coefficients in the decomposition of equation~(\ref{forged_state}), then sampling from them is a simple task that can be achieved in time linear with the number of coefficients~\cite{eddinsDoublingSizeQuantum2022}. However if the number of coefficients considered grows fast with the system size, naive sampling can become prohibitive. On the other hand, the use of a generative model can open the possibility of sampling distributions of up to exponentially many coefficients. Artificial neural networks are likely the most powerful known computational technique to approximate high-dimensional probability densities. 

\begin{figure}[t]
\centering
  \includegraphics[width=\linewidth]{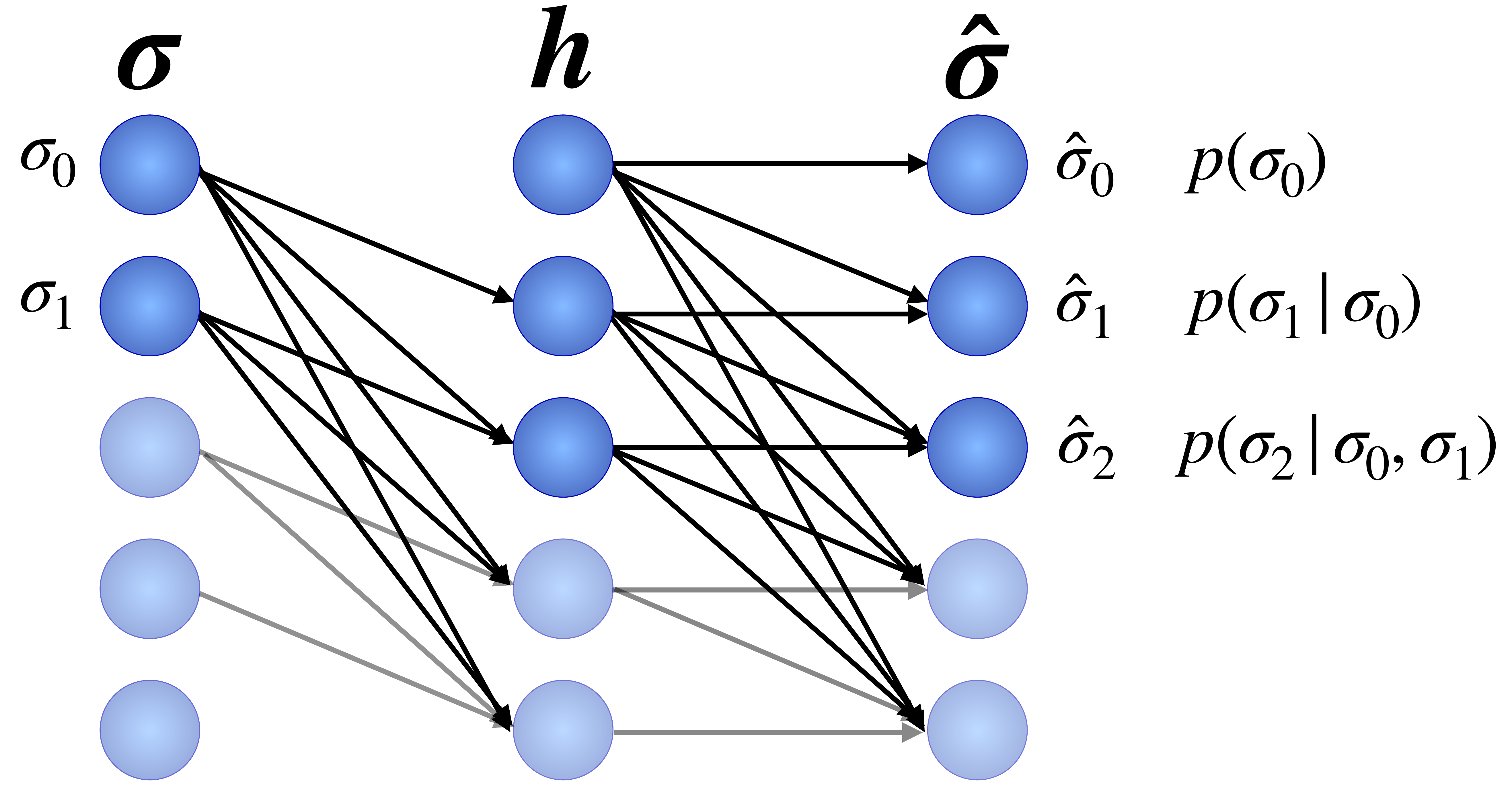}
  \caption{Recurrent neural network with one hidden layer. The first input is sampled directly from the neural network via 
  $p(\sigma_0) = (1 - \hat{\sigma}_0)^{\hat{\sigma_0}}$ and does not depend on the input of the NN. The conditional probability to sample $\sigma_1$ depends then on $\sigma_0$ which we feed as an input to the NN and $p(\sigma_1 | \sigma_0) = (1 - \hat{\sigma_1})^{\hat{\sigma_1}}$}.
  \label{fig:ARNN}
\end{figure}

To see how they can play a role here, we use a parametrized function $\lambda_{\Btheta}(\Bsigma)$ to approximate $\lambda_{\Bsigma}$ with trainable parameters $\Btheta$. The Schmidt coefficients are normalized $\sum_{\Bsigma} |\lambda_{\Bsigma}| ^2 = 1$ which allows us to interpret $\vert \lambda_{\Bsigma}\vert ^2$ as a probability density over the spin variables.  We use then neural networks representations for this probability density. We specifically consider 
auto-regressive neural networks (ARNNs)~\cite{oordConditionalImageGeneration2016a} to model these probabilities. The coefficients $\lambda_{\Bsigma}$ can be chosen to be real or complex. For our experiments here we assumed them to be real and non-negative. The main requirements on the models is that we can sample them efficiently and that we have access to the function $R(\Bsigma, \Bsigma') = \lambda_{\Bsigma'} / \lambda_{\Bsigma}$ to evaluate the classical part of equation~\ref{eq:mu_alpha_beta}. ARNNs are particularly suitable for the task at hand because they can be sampled directly, and efficiently, without Markov chain Monte Carlo. Also, they allow to efficiently compute normalized probability densities, hence the ratio $R(\Bsigma,\Bsigma^{\prime})$.
The idea of an ARNN is to model the probability distribution $p(\Bsigma)$ of a sample $\Bsigma$ with the conditional probabilities $p(\Bsigma) = \prod_i p(\sigma_i \vert \Bsigma_{<i})$, where $\Bsigma_{<i}$ denotes all the bits before $\sigma_i$. To achieve this we use a dense ARNN architecture shown in Figure~\ref{fig:ARNN}. The ARNN returns an output $\hat{\Bsigma} \in [0,1]^N$ for each input $\Bsigma \in \{0,1 \}^N$ and the probability $p(\sigma_i = 1)$ is given by $\hat{\sigma}_i$. 
To obtain a sample from the ARNN one can sample $\Bsigma$ elementwise from the Bernoulli distribution $(1 - \hat{\sigma}_i)^{\hat{\sigma}_i}$ starting at the element $\sigma_0$ and recursively sampling all elements $\sigma_i$, given an input to the ARNN $\sigma_{<i}$.
\subsection{Quantum Circuit Ansatz}
\begin{figure}[t]
\centering
  \includegraphics[width=\linewidth]{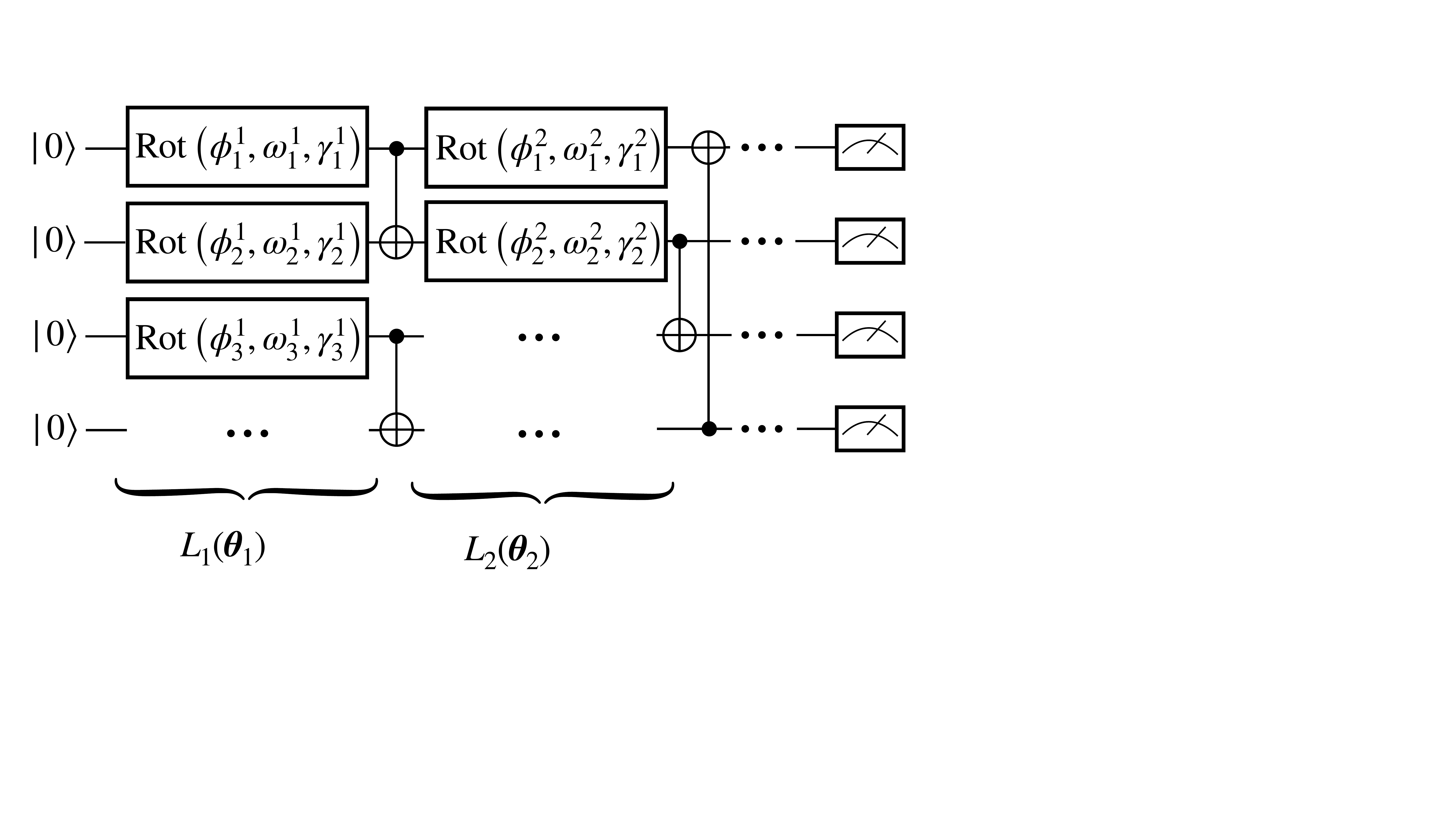}
  \caption{Hardware efficient ansatz for 4 qubits, where each layer $L_j$ consists of general SU(2) rotations acting on all qubits with each 3 parameters. The rotational gates are followed by CNOT gates arranged like a brick wall.}
  \label{fig:Circuit}
\end{figure}

For our hybrid quantum-classical scheme, one needs to pick a parametrized quantum circuit to be used as ansatz. For the systems considered in this work we parameterize the unitary $U$ as shown in Figure~\ref{fig:Circuit}, using a repeating pattern of general one-qubit $SU(2)$ rotations $\text{Rot}(\phi, \omega, \gamma) = R_z(\gamma) R_y(\omega) R_Z(\phi)$ followed by $\text{CNOT}_{i, i+1}$ gates acting on neighbouring qubits $i$ and $i+1$. We refer to the combination of $\text{Rot}$ gates followed by CNOTs as a layer $L_j(\boldsymbol{\theta_j})$ of this ansatz, with $\boldsymbol{\theta_j}$ representing all the parameters of the layer $j$. In even layers the control of the CNOT gate $i$ acts on the even qubits and in odd layers the control acts on odd qubits. We also assume periodicity $(i+1)\mathrm{mod}(N)$. This ansatz is often referred to as a hardware efficient ansatz.
\subsection{Energy minimization}
In the following, we write the dependence of the NN model on their parameters explicitly $\lambda_{\Bsigma} \rightarrow \lambda_{\Btheta}(\Bsigma)$. Furthermore, we introduce the short-hand notation $\braket{ \Bsigma \vert U_{\Bomega}^{\dag} 
    \mathcal{O} U_{\Bomega} \vert \Bsigma } = \braket{\mathcal{O}}_{\Bsigma}^{\Bomega}$.
The energy of a general system is given by
\begin{align}
    E(\Btheta, \Bomega) = \braket{H_A} + \braket{H_B} + \sum_{\substack{i \in A \\ j\in B}} \braket{O_i \otimes O_j},
\end{align}
which is to be minimized via gradient descent. The sum over $i \in A$ and $j \in B$ indicates all the terms of the system acting on both subsystems $A$ and $B$ where the index $i$ is in subsystem $A$ and $j$ is in subsystem $B$.  To take the gradient with respect to the parameters $\Btheta$ and $\Bomega$ we make use of the log likelihood ``trick'' which allows us to rewrite 
\begin{align}
    \nabla_{\Btheta} \sum_{\Bsigma} p_{\Btheta}(\Bsigma) f({\Bsigma}) = \mathds{E} \big[ \nabla_{\Btheta} \log p_{\Btheta}(\Bsigma) f(\Bsigma) \big].
\end{align}
Where the expectation value $\mathds{E}\big[ \cdot \big]$ can be estimated with samples from the distribution $p_{\Btheta}(\Bsigma)$, i.e. one can sum $\sum_{\Bsigma \sim p_{\Btheta}(\Bsigma)}$.
Therefore, the derivative of the energy with respect to the parameters of the NN $\Btheta$ reads:
\begin{align}
    &\nabla_{\Btheta} E(\Btheta, \Bomega)
    =  \mathds{E}_{\lambda}
      \Bigg[ \nabla_{\Btheta} \log(\Plambda^2) \Bigg(
    \braket{H_A}_{\Bsigma}^{\Bomega} + \braket{H_B}_{\Bsigma}^{\Bomega} \nonumber \\
    &+ \sum_{\substack{O_i \in A \\ O_j\in B}} \Bigg\{
    \braket{O_i O_j}_{\Bsigma}^{\Bomega}
      + \sum_{\alpha, \beta \in \{0,1\}}  \mathds{E}_{p} \big[ R(\Bsigma, \Bsigma') \big]
      \Bigg\} \Bigg)  \nonumber \\
      & + 
       \sum_{\substack{O_i \in A \\ O_j\in B}} 
      \sum_{\alpha, \beta \in \{0,1\}}
      \mathds{E}_{p} \big[ \nabla_{\Btheta} R(\Bsigma, \Bsigma') \big] \Bigg].
\end{align}
The two expectation values are according to the distributions $\mathds{E}_{\lambda} = \mathds{E}_{\Bsigma \sim \Plambda^2}$ and
$\mathds{E}_p = \mathds{E}_{\Bsigma' \sim p_{\alpha, \beta}^{i,j}(\Bsigma' | \Bsigma)}$. We here used the short hand notation $p_{\alpha, \beta}^{i,j} (\Bsigma' | \Bsigma) =  \vert \braket{\Bsigma' \vert U^{\dag} C_{\alpha, \beta}^{i,j} U\vert \Bsigma } \vert^2$ to indicate that the Clifford $C_{\alpha, \beta}^{i,j}$ depends on $\alpha$ and $\beta$ and also on the operators $O_i$ and $O_j$.
The derivative of the energy with respect to the parameters $\Bomega$ of the unitaries is:
\begin{align}
    &\nabla_{\Bomega} E(\Btheta, \Bomega) =  \mathds{E}_{\lambda} \Bigg[   \nabla_{\Bomega} 
    \braket{H_A}_{\Bsigma}^{\Bomega} + 
    \nabla_{\Bomega}  \braket{H_B}_{\Bsigma}^{\Bomega} \nonumber \\
    &+ \sum_{\substack{O_i \in A \\ O_j\in B}} \Bigg\{ \nabla_{\Bomega} 
    \braket{O_i O_j}_{\Bsigma}^{\Bomega} \nonumber \\
      &+ \sum_{\alpha, \beta \in \{0,1\}}
      \mathds{E}_p \big[ \nabla_{\Bomega} \log p_{\alpha, \beta}^{i,j}(\Bsigma' | \Bsigma)  R(\Bsigma, \Bsigma') \big] \Bigg\} \Bigg].
\end{align}
To make the energy minimization of the parameterized quantum circuit scalable and therefore, independent of the exact evaluation of $p_{\alpha, \beta}^{i,j}$ on can use gradient free methods, such as for example simultaneous perturbation stochastic approximation (SPSA)~\cite{spall1998overview}.

\section{Spins in one dimension}
For a better understanding of the algorithm we present the 1D TFIM model with PBC in detail. We have two equally sized subsystems $A$ and $B$ with $N$ qubits each. The TFIM Hamiltonian can be split into three operators which we can evaluate separately. The operators $H_A$ and $H_B$ act only on subsystem $A$ or $B$ and $\sum_i O_A^i \otimes O_B^i$ connects the two subsystems. More concretely, $H_A = H_B = \sum_{i}^{N-1} Z^i Z^{i+1} + \sum_i^N X^i$ and $\sum_i O_A^i \otimes O_B^i = Z_A^N \otimes Z_B^1 + Z_A^1 \otimes Z_B^N$ because of the periodic boundary conditions. The variational energy of the system is given by
\begin{align}
    E(\Btheta, \Bomega) = 2 \braket{H_A} + 2 \braket{Z_A^1 \otimes Z_B^N},
\end{align}
which is to be minimized via gradient descent. Note that $\braket{Z_A^1 \otimes Z_B^N} = \braket{Z_A^N \otimes Z_B^1}$ because of the translation invariance. Hence, the factor 2.
\begin{figure}[t]
\centering
  \includegraphics[width=\linewidth]{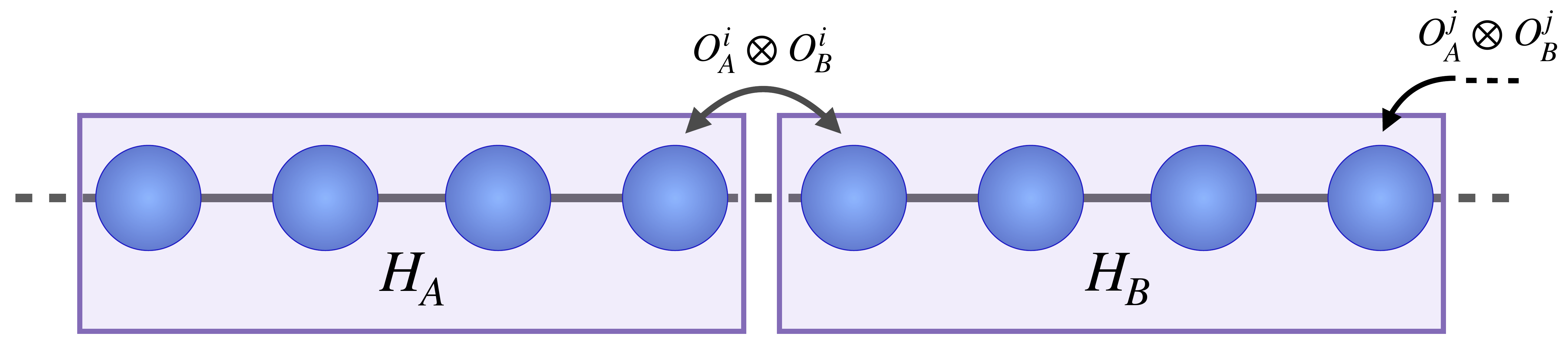}
  \caption{1D transverse field Ising (TFIM) Hamiltonian with 8 qubits and periodic boundaries.}
  \label{fig:1DTFIM}
\end{figure}
The Hamiltonians $H_A$ and $H_B$ act only on one of the subsystems and, therefore, one can evaluate the energy directly via 
\begin{align}
    \braket{H_A} = \sum_{\Bsigma \sim \lambda_{\Bsigma}^2} \braket{ \Bsigma \vert U^{\dag} H_A U \vert \Bsigma }.
\end{align}
One can sample $\Bsigma$ from the classical distribution $\lambda_{\Bsigma}^2$ and calculate the mean of the expectation values $\braket{ \Bsigma \vert U^{\dag} H_A U \vert \Bsigma }$. 

Analogously, one can evaluate the first term of equation~\ref{eq:Expect_off_diagonal} for operators that act on both systems, e.g. $\braket{O_A \otimes O_B}$ simply by replacing $\braket{ \Bsigma \vert U^{\dag} H_A U \vert \Bsigma }$ with $\braket{\Bsigma \vert U^{\dag} O_AO_B U \vert \Bsigma }$. To be even more specific let's assume we evaluate the expectation value $\braket{Z_A^N \otimes Z_B^1}$. In this case the first term of equation~\ref{eq:Expect_off_diagonal} reduces to evaluating the expectation value $\braket{\Bsigma \vert U^{\dag} Z^1Z^N U \vert \Bsigma }$. 
For the second term of equation~\ref{eq:Expect_off_diagonal}, $\mu_{\alpha, \beta}$ has to be calculated. To do so we first sample $\Bsigma \sim \lambda_{\Bsigma}^2$, prepare the circuit $U^{\dag} C_{\alpha, \beta} U \ket{\Bsigma}$ and measure each qubit in the $Z$ basis to obtain samples of $\Bsigma'$. Then we evaluate $R(\Bsigma, \Bsigma')$ and take the average over several samples of $\Bsigma'$ to approximate $\sum_{\Bsigma'} R(\Bsigma, \Bsigma') p(\Bsigma' \vert \Bsigma)$. We repeat this procedure for many samples $\Bsigma \sim \lambda_{\Bsigma}^2$ and take the average. The coefficients are $a_0 = a_{0,0} = a_{1,1} = 1$ and $a_{0,1} = a_{1,0} = -1$ for the operators $Z_A^i \otimes Z_B^j$.

\begin{figure}[t!]
\centering
  \includegraphics[width=\linewidth]{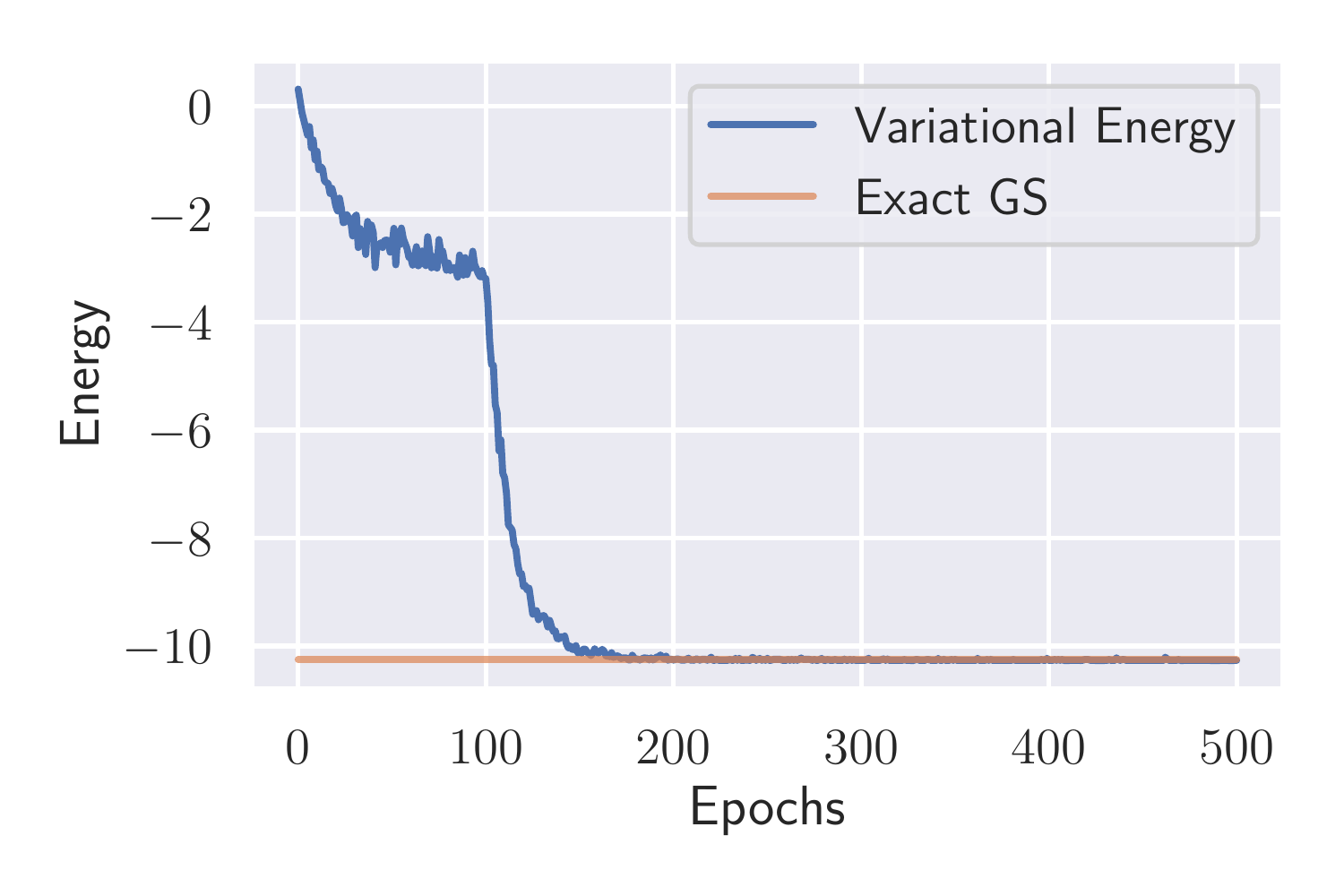}
  \caption{\textbf{1D TFIM}: Convergence of the energy 8 qubit forged state in 1D. The first 100 epochs we only optimize the parameters of the quantum circuit $\Bomega$ until convergence and then we start optimizing as well the parameters of the neural network $\Btheta$.}
  \label{fig:1DTFIM_energy}
\end{figure}

\begin{figure}[t]
\centering
  \includegraphics[width=\linewidth]{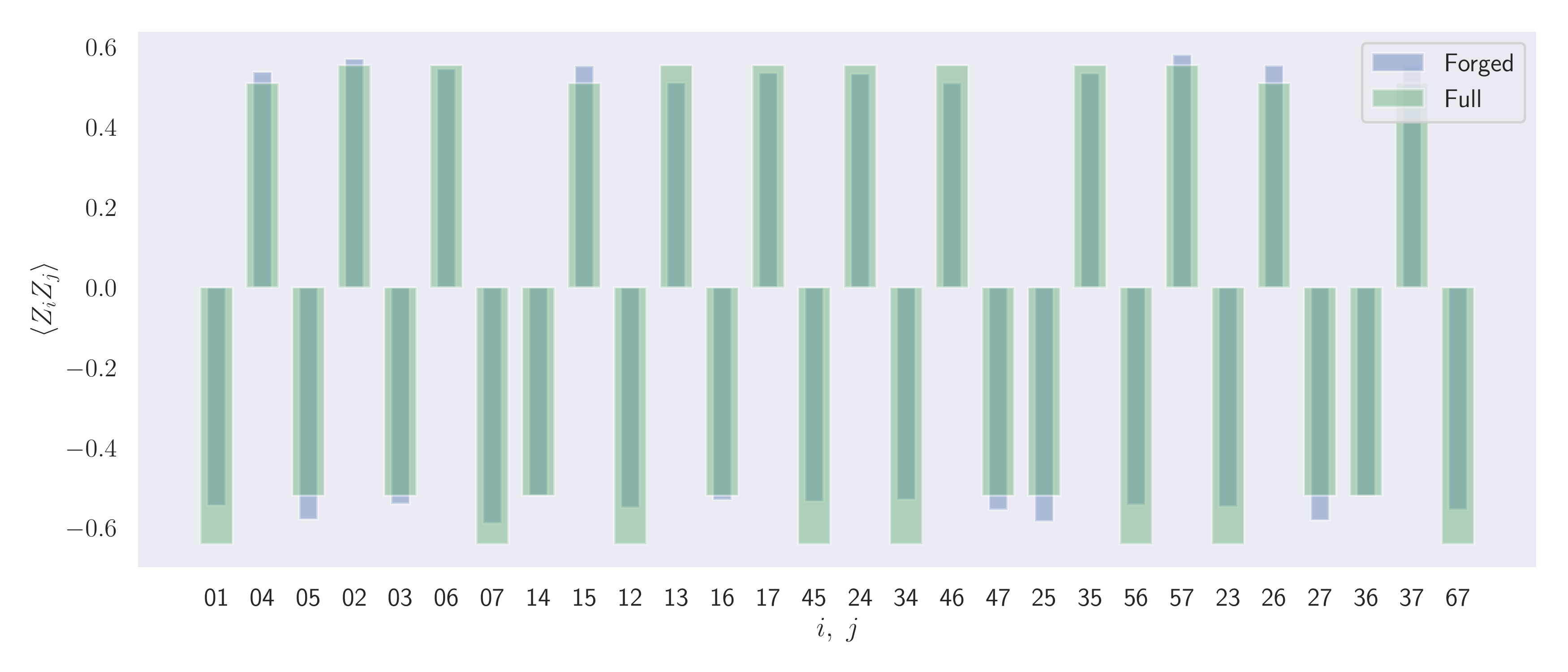}
  \caption{Correlators 1D: $\langle Z_i Z_j \rangle$ expectation values of the 8 qubit forged state in 1D. The forged correlators are in blue, the correlators of the exact calulcation of the full system are in green.}
  \label{fig:1DTFIM_correlator}
\end{figure}

In Figure~\ref{fig:1DTFIM_energy} we show the gradient descent progress of the energy optimization of a one dimensional TFIM Hamiltonian at the critical point with 8 spins. We find the ground state energy with high accuracy and we can reproduce the spin-spin correlators $\braket{Z_i Z_j}$ of the ground state shown in figure~\ref{fig:1DTFIM_correlator}. 
\section{Spins in two dimensions}
\begin{figure}[t]
\centering
  \includegraphics[width=\linewidth]{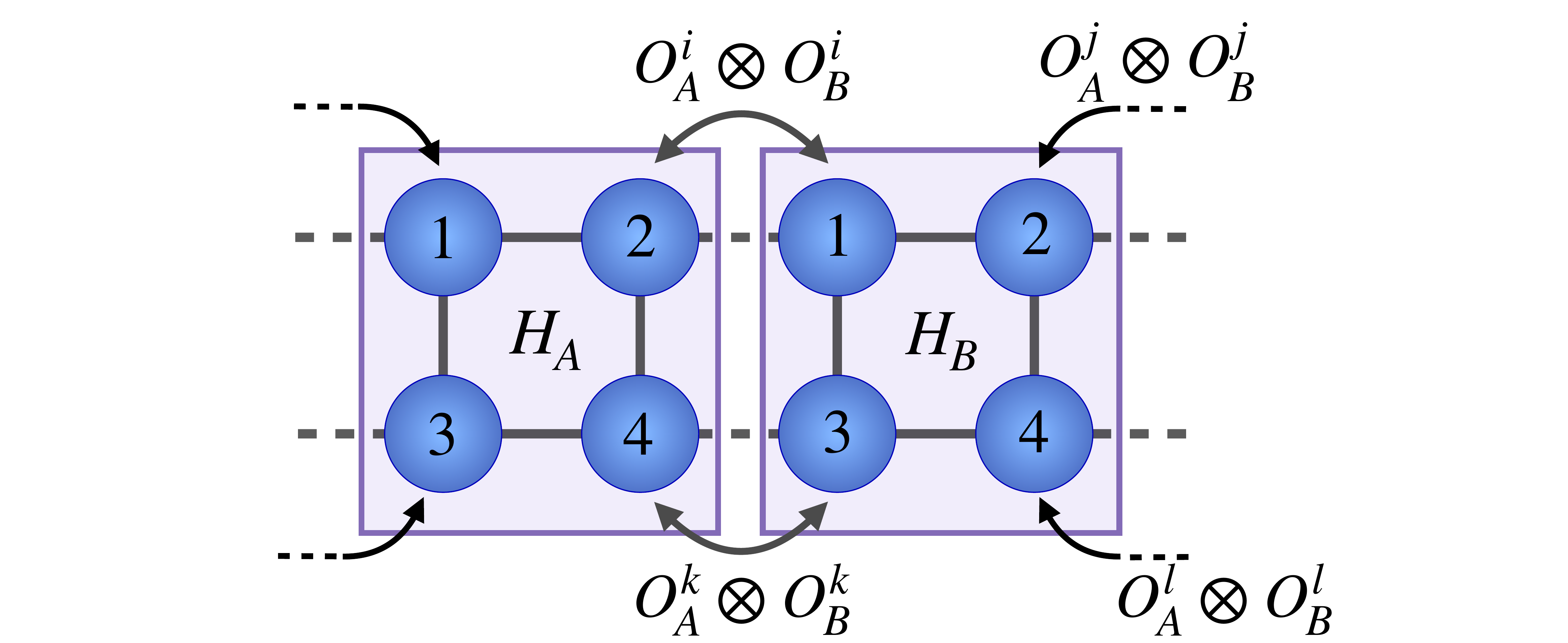}
  \caption{2D transverse field Ising (TFIM) Hamiltonian with $2 \times 4$ qubits with peridoic boundaries in the horizontal direction.}
  \label{fig:2DTFIM_Hamiltonian}
\end{figure}
\begin{figure}[t]
\centering
  \includegraphics[width=\linewidth]{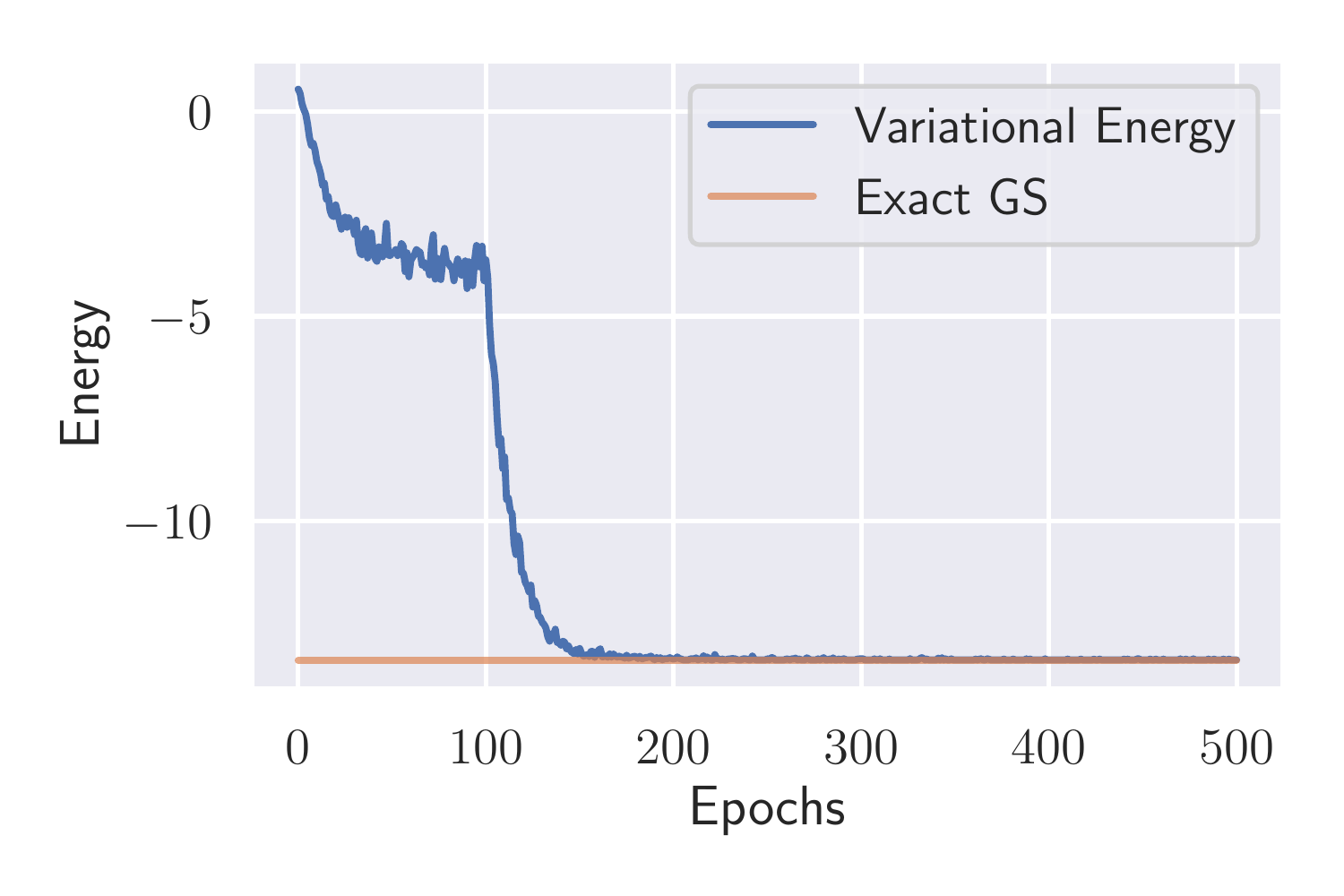}
  \caption{\textbf{2D TFIM}: Convergence of the energy of a 8 qubit forged state in 2D. The first 100 epochs we only optimize the parameters of the quantum circuit $\Bomega$ until convergence and then we start optimizing as well the parameters of the neural network, $\Btheta$.}
  \label{fig:2DTFIM_energy}
\end{figure}
\begin{figure}[t]
\centering
  \includegraphics[width=\linewidth]{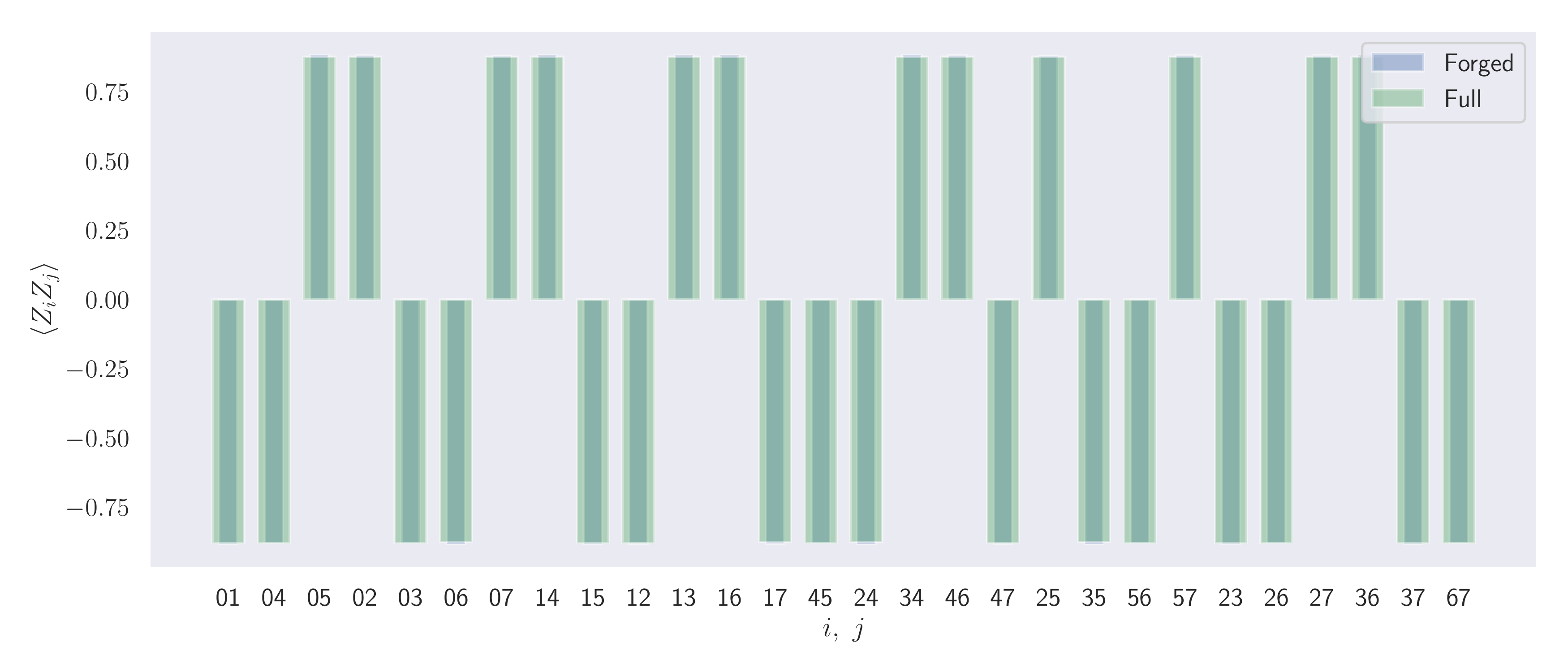}
  \caption{Correlators 2D: $\langle Z_i Z_j \rangle$ expectation values of the 8 qubit forged state on a 2D 2x4 grid. The forged correlators are in blue, the correlators of the exact calulcation of the full system are in green.}
  \label{fig:2DTFIM_correlator}
\end{figure}

For the 2 dimensional case, we study the TFIM Hmailtonian as shown in Figure~\ref{fig:2DTFIM_Hamiltonian}. Compared to the one dimensional case we add two more terms that couple the subsystems. Therefore, the energy is
\begin{align}
    E(\Btheta, \Bomega) =  2 \braket{H_A} +2 \braket{Z_A^1 \otimes Z_B^2} + 2 \braket{Z_A^3 \otimes Z_B^4},
\end{align}
with $H_A = H_B = \sum_{\langle i,j \rangle} Z_i Z_j + \sum_i X_i$. Again, we used the fact that for our ansatz where the unitary $U_{\Bomega}$ for both subsystems is equal if follows that $\braket{Z_A^i \otimes Z_B^j} = \braket{Z_A^j \otimes Z_B^i}$. The optimization of the energy is equivalent to the 1D case and the convergence of the energy is shown in Figure~\ref{fig:2DTFIM_energy}. The energy converges to the value calculated by exact diagonalization and the qubit-qubit correlators, shown in figure~\ref{fig:2DTFIM_correlator}, coincide highly with the exact values.
\section{Lattice Fermions}
\begin{figure}[t]
\centering
  \includegraphics[width=\linewidth]{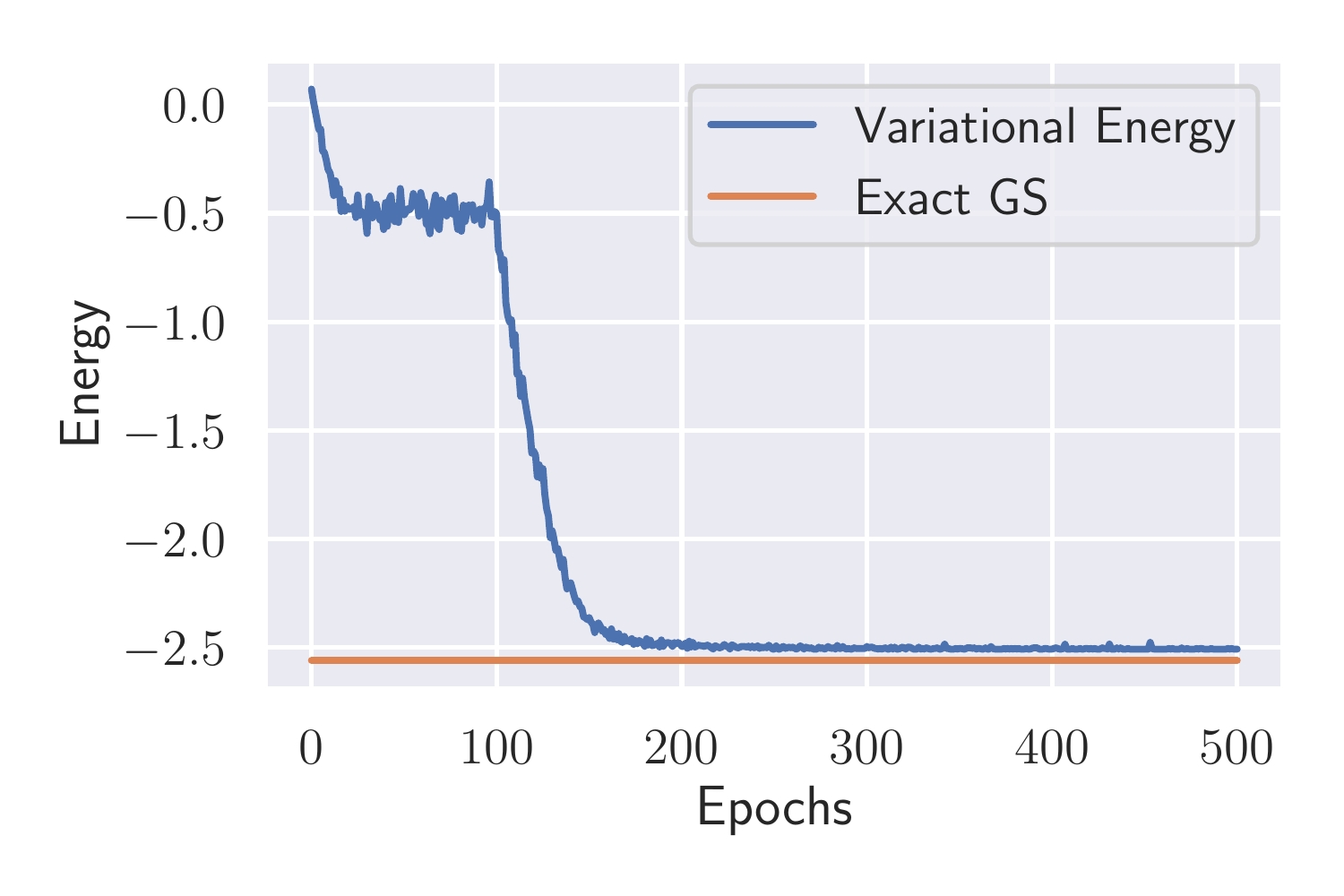}
  \caption{\textbf{Fermionic system}: Convergence of the energy of a 2x2 spinless fermions forged state. The first 100 epochs we only optimize the parameters of the quantum circuit $\Bomega$ until convergence and then we start optimizing as well the parameters of the neural network,  $\Btheta$.}
  \label{fig:Fermi_energy}
\end{figure}

As long as the qubit Hamiltonian has permutational symmetry along the partition of the system, our forging procedure can also be applied to fermionic systems. With the help of the Jordan-Wigner transformation we can map the t-V-Hamiltonian 
\begin{align}
     H = - t \sum_{\langle i, j \rangle} (a^\dagger_i a_j + a^\dagger_j a_i)
        + V \sum_{\langle i, j \rangle} a^\dagger_i a_i a^\dagger_j a_j
\end{align}
to a qubit Hamiltonian. For example, for a $2 \times 2$ system of spinless fermions with periodic boundaries and $t = V =1$ the qubit Hamiltonian reads
\begin{align}
H_{qb} &= \frac{1}{2} \big[ 
 X_1 X_2 +
X_1 Z_2 X_3 +
Y_1 Y_2 +
Y_1 Z_2 Y_3   \\ \nonumber
 & + X_2 Z_3 X_4 
 + Y_2 Z_3 Y_4 
 + X_3 X_4 
 + Y_3 Y_4 \big] \\ \nonumber
&+\frac{1}{4} 
 \big[Z_1 Z_2 +
Z_1 Z_3 +
Z_2 Z_4 +
Z_3 Z_4 \big].
\end{align}
This Hamiltonian can be split into two partitions $H_A$ and $H_B$ and an interacting term $H_I$ such that $H_{qb} = H_A + H_B + H_I$ and $H_A = H_B$. Therefore the mirror symmetry of the system is fullfilled and our ansatz can capture the groundstate of this system.
We split the 4-qubit system into two subsystems, where qubit 1 and 2 build subsystem A and qubit 3 and 4 build subsystem B. As for the previous examples, the observables that act only on one subsystem are evaluated with $\mu = \sum_{\Bsigma} \lambda_{\Bsigma}^2 \braket{\Bsigma \vert U^{\dag} O_A U \vert \Bsigma}$. For the observables that act on both subsystems we obtain the expectation value through equation~\ref{eq:Expect_off_diagonal}. The unitary $C_{\alpha, \beta}$ for general observables $O_A$ and $O_B$ is given by $C_{\alpha, \beta} = \frac{1}{2}(I + (-1)^{\alpha}O_A + (-1)^{\beta}O_B - (-1)^{\alpha+\beta} O_A O_B )$. More details and how to decompose them into standard qubit gates are described in~\cite{eddinsDoublingSizeQuantum2022}. In figure \ref{fig:Fermi_energy} we show the energy throughout the training for spinless fermions.
\section{Methods}
The numerical simulations of the quantum circuits have been done in pennylane~\cite{bergholmPennyLaneAutomaticDifferentiation2018} with the JAX~\cite{jax2018github} backend. The optimization of the classical part has been performed in Netket\cite{vicentiniNetKetMachineLearning2021}. All the code is accessible on Github~\cite{Huembeli_Github_2022}.

\section{Discussion and Conclusion}
Through the combination of classical probabilistic models and quantum circuits we have demonstrated expressive variational quantum-classical ans\"atze with an overall reduced amount of computational resources. 
We have proposed and numerically tested a quantum-classical entanglement forging approach with an overall polynomial cost in the estimation of observables. An auto-regressive neural networks was used to learn the Schmidt decomposition of a target state to approximate it. We have shown through proof-of-principle numerics that the method works on 1D and 2D transverse field Ising model, finding ground states energies and two-point correlation functions with high accuracy. Furthermore, we have shown that our method can also be applied to fermionic systems in presence of permutational symmetries of subsystems. It remains not clear whether the approach here presented can address systems that do not have subsystem permutational symmetries. We foresee this could be an interesting research question.   
\section{Acknowledgements}
We thank Sergey Bravyi for insightful discussions.
P.H. acknowledges the use of IBM Quantum services, and advanced services and support provided by the IBM Quantum Researchers Program

\bibliographystyle{apsrev4-1.bst} 
\bibliography{bibliography}

\end{document}